\documentclass[aps,epsfigure,twocolumn,superscriptaddress,showkeys,nofootinbib]{revtex4-1}
\usepackage[colorlinks=true,linkcolor=blue,urlcolor=blue,citecolor=blue,pdfusetitle]{hyperref}
\usepackage[utf8]{inputenc}
\usepackage[english]{babel}
\usepackage{amsmath}
\usepackage[caption = false]{subfig}

\usepackage{graphicx,epstopdf}
\usepackage{blindtext}
\usepackage[table,xcdraw]{xcolor}
\usepackage{lipsum}
\usepackage{amsfonts}
\usepackage{bbm}
\usepackage{amssymb}
\usepackage{enumerate}
\usepackage{color}
\usepackage{latexsym}
\usepackage{times,txfonts}
\usepackage[normalem]{ulem}
\graphicspath{ {./figures/} }

\usepackage{amsmath}
\usepackage{lipsum}

\begin{document}

\title{Evolutionary quantum feature selection}

\author{Anton Simen Albino}
\email{anton.simen@kipu-quantum.com}
\affiliation{Kipu Quantum, Berlin, Germany.}
\affiliation{Latin American Quantum Computing Center, SENAI CIMATEC, Salvador, Brazil.}

\author{Otto Pires}
\email{otto.pires@fbter.org.br}
\affiliation{Latin American Quantum Computing Center, SENAI CIMATEC, Salvador, Brazil.}

\author{Mauro Nooblath}
\email{mauro.neto@fbter.org.br}
\affiliation{Latin American Quantum Computing Center, SENAI CIMATEC, Salvador, Brazil.}

\author{Erick Giovani Sperandio Nascimento}
\email{erick.sperandio@surrey.ac.uk}
\affiliation{Surrey Institute for People-Centred Artificial Intelligence, University of Surrey, Guildford, United Kingdom}



\begin{abstract}

Effective feature selection is essential for enhancing the performance of artificial intelligence models. It involves identifying feature combinations that optimize a given metric, but this is a challenging task due to the problem's exponential time complexity. In this study, we present an innovative heuristic called Evolutionary Quantum Feature Selection (EQFS) that employs the Quantum Circuit Evolution (QCE) algorithm. Our approach harnesses the unique capabilities of QCE, which utilizes shallow depth circuits to generate sparse probability distributions. Our computational experiments demonstrate that EQFS can identify good feature combinations with quadratic scaling in the number of features. To evaluate EQFS's performance, we counted the number of times a given classical model assesses the cost function for a specific metric, as a function of the number of generations.

\end{abstract}
\maketitle

\section{Introduction}

Quantum Feature Selection (QFS) is a novel approach to Feature Selection (FS) in Machine Learning (ML) that leverages principles of Quantum Computing (QC) to enhance the efficiency and effectiveness of traditional FS methods. The most informative features are typically selected in traditional FS methods based on their correlation with the target variable or their predictive power. However, these methods can struggle with high-dimensional datasets, a phenomenon known as the curse of dimensionality \cite{Mcke2023}. On the other hand, Evolutionary Algorithms (EAs) are a family of optimization algorithms that are inspired by the process of natural selection and evolution. These algorithms use a population of candidate solutions and iteratively improve them over generations through selection, recombination, and mutation operations. \cite{yu2010introduction}\cite{julian2022}. 

The rapid increase in the amount of data has made it challenging to keep up with the computational demands of traditional FS methods \cite{electronics11193177}. As a result,  researchers have explored alternative perspectives such as quantum computing and evolutionary algorithms. A procedure was described for a novel FS algorithm based on a Quadratic Unconstrained Binary Optimization (QUBO) problem for reducing model complexity in Machine Learning \cite{Mcke2023}. The algorithms selects a specific number of features based on their importance and redundancy, and the direct approach used in the algorithm yields higher quality solutions compared to iterative or greddy methods. The QUBO problems are particularly interesting because they can be solved on Quantum Hardware, which is why the proposed algorithm was evaluated using a classical computer, a quantum gate computer, and a quantum annealer. The proposed FS algorithm based on QUBO is a promising approach to address the challenges posed by the growing amount of data in Machine Learning.

 Other study was realized by \cite{zoufal2023variational} that describes a variational quantum algorithm designed to solve unscontrained black box binary optimization problems, where the objective function is given as a black box. Unlike typical algorithms for optimization where a classical objetive function is provided as a Quandratic Uncontrained Binary Optimization problem and mapped toa sum of Pauli operators, this algorithm directly handles the black box objective function. The algorithm´s theorical justification is based on convergence guarantees of quantum imaginary time evolution. The authors demonstrated that the quantum method produced competitive, and in certain aspects, even better perfomance compared to traditional FS techniques used in today´s industry. This suggests that quantum algorithms could potentially offer significant advantages over classical methods in FS and other optimization problems. However, further research is necessary to explore the full capabilities of this approach and it´s potential applications in real-world scenarios. 

This paper focuses on the challenge of effective FS for artificial intelligence models due to exponential time complexity. To address this challenge, we propose an innovative heuristic called Evolutionary Quantum Feature Selection (EQFS) that uses the Quantum Circuit Evolution (QCE) algorithm. The QCE uses shalllow depth circuits to sparse probability distributions, because of this they can be useful to be applied in Noisy Intermediate-Scale Quantum(NISQ) devices, which EQFS harnesses to identify good feature combinations with quadratic scaling in the number of features. We evaluated EQFS´s perfomance by counting the number of times a given classical model assesses the cost function for a specific metric as a function  of the number of generations. This work was organized as follows: In part \ref{II}, a brief description of the model used was made. In part \ref{III}, the results were discussed and finally in part \ref{IV} the conclusions of this work.




\section{Quantum Feature Selection} \label{II}

\begin{figure}[h!]
    \centering
    \includegraphics[width=7 cm]{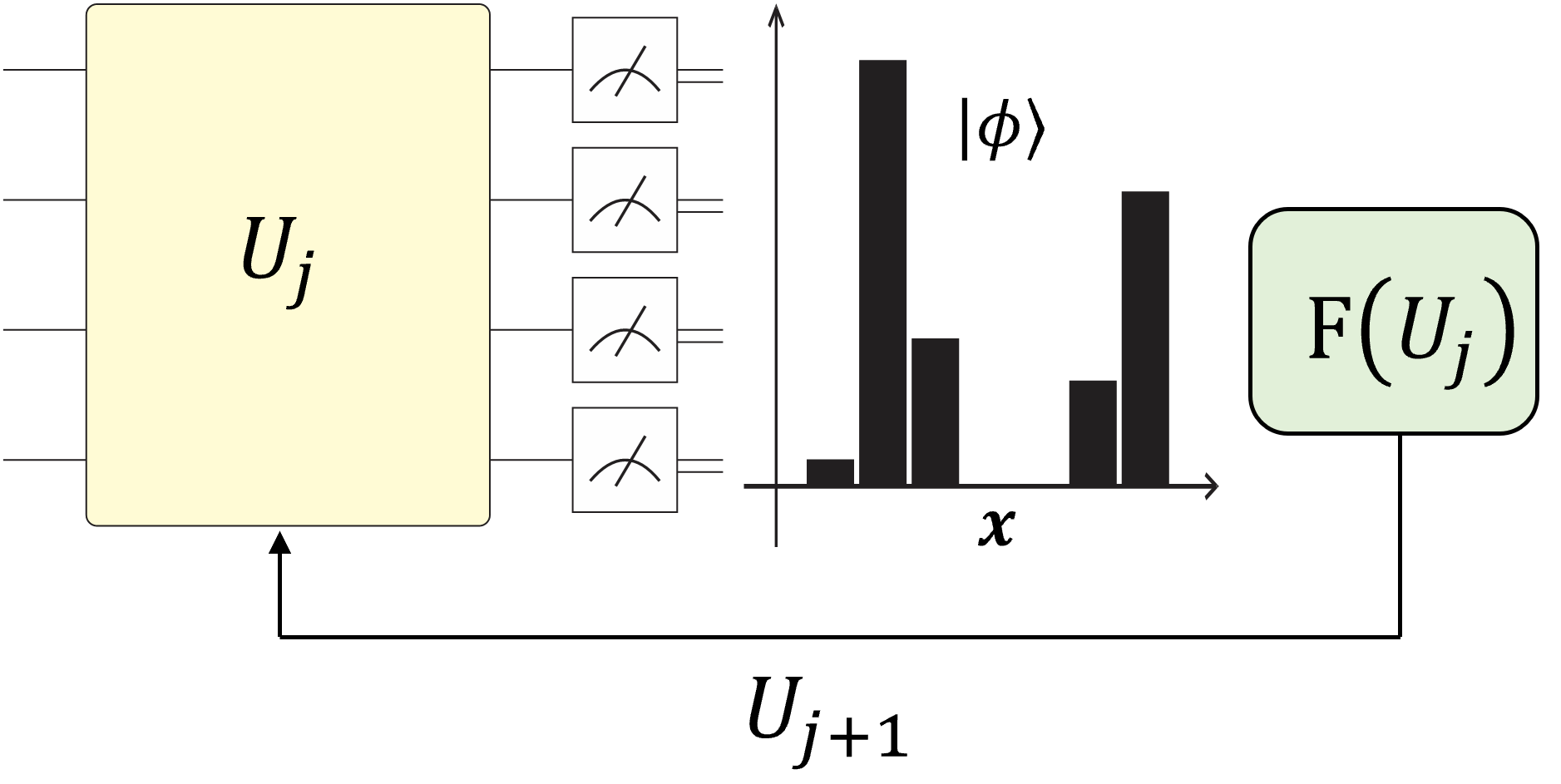}
    \caption{Evolutionary procedure where a quantum circuit $U$ evolves over the generations in order to minimize a target function given by $F(U)$ evaluated from the quasi-probability distribution $|\phi\rangle$.}
    \label{fig:qfs}
\end{figure}

The procedure that will be described to perform QFS uses a hybrid approach, where a quantum evolutionary algorithm plays the role of feature combination optimizer and works together with a classical algorithm that evaluates feature combinations in a supervised learning model. Let $X$ be a dataset of dimensionality $\text{dim}(X) = n$. Each sample of $X$ can be represented as an $n-$dimensional vector, $\mathbf{v}$ with its associated vector of binary values, $\mathbf{x} = \left(x_0, x_1, x_2, . .., x_n\right)$, which plays the role of indicating whether a variable will feed ($x_i = 1$) or not ($x_i=0$) the classical model. Let a metric function $f(\mathbf{x})$ evaluate the model quality given the $\mathbf{x}$ combination of features. Let an initial quantum state be given by $|\psi\rangle = |0\rangle ^{\otimes n}$ and the unit transformation, $U|\psi\rangle = |\phi\rangle$ being $U$ generated by a quantum circuit that can be subjected to mutations over the generations (see Fig. \ref{fig:qfs}).  The QFS objective function can be writen as 

\begin{equation}
    F(U) = \sum_{\mathbf{x}}|\langle\mathbf{x}|\phi\rangle|^2 f(\mathbf{x}).
\end{equation}

It is important to note that here we are considering that $|\phi\rangle$ is not a proper quantum state, but rather a vector of quasi-probabilities after a polynomial set of measurements on $U|\psi\rangle$. Since we have $\mathcal{O}\left(2^n\right)$ possible solutions to the problem, a number of measures $ m = \mathcal{O}\left(\text{poly}(n)\right)$ ensures that the approximate solution is found with a time complexity that scales polynomially with the number of variables.

In this work, the algorithm used to evolve $U$ is Quantum Circuit Evolution, proposed by \cite{qce}. At each generation, $\lambda$ copies of $U$ are created and a mutation operation with a respective probability is applied to each generation. Possible mutations are \textbf{insert} a new gate; \textbf{modify} a rotation angle of a single or two-qubit gate on the current circuit; \textbf{delete} one of the current gates and \textbf{swap} a two-qubit gate (flip target and control). Each of these mutations has its respective probability of occurring on each of the $\lambda$ copies of $U$ and the $\mu$ best individuals are carried over to the next generation, characterizing an elitist procedure known in the literature as ($\mu + \lambda$)EA.

In order to estimate the number of times the objective function is evaluated, consider $\Omega$ to be the set of probability amplitudes derived from $|\phi\rangle$. If we take $m$ as the number of measurements performed on the quantum circuit, we have $\text{dim} \ \Omega \geq m$. However, because it is a heuristic whose initial and final generations have sparse states - since the quantum circuit starts with a small depth - we can consider that very possibly $\text{dim} \ \Omega \gg m$. Given a number $K$ of generations and $\lambda$ copies, we empirically observe that, for a small $k$, the total number of model evaluations in each generation, $ \sum_{i=0}^{\lambda} \text{dim} \Omega_i $, can be approximated by a linear function with dependence on $k$ and with a fixed constant defined for $m$ as
\begin{equation}
     \sum_{i=0}^{\lambda} \text{dim} \Omega_i = \frac{m}{K} k.
\end{equation}
Therefore, we can approximate the number $t$ of times that the objective function, $f(\textbf{x})$, is evaluated by calculation the Area Under the Curve (AUC), given by

\begin{equation}
     t \approx \int_{0}^{K} \frac{m}{K} k \,dk.
\end{equation}

Note that to approximate the solution in reasonable time, we can choose $m$ and $K$ appropriately. The experiments showed that for $m$ and $K$ being $\mathcal{O}(\text{linear}(n))$, the heuristic can already find better solutions than for $\mathbf{x}$ containing all features. 

\section{Results and Discussion}\label{III}

The results of the proposed feature selection procedure using quantum computing are presented and analyzed in this section. This method aims to address the challenges faced by classical feature selection algorithms in handling high dimensional datasets. The procedure is designed to improve the accuracy and efficiency of feature selection. Experimental results are presented to demonstrate the effectiveness of the proposed method and its comparison to classical methods.

\begin{figure}[h!]
    \centering    \includegraphics[width=7.5 cm]{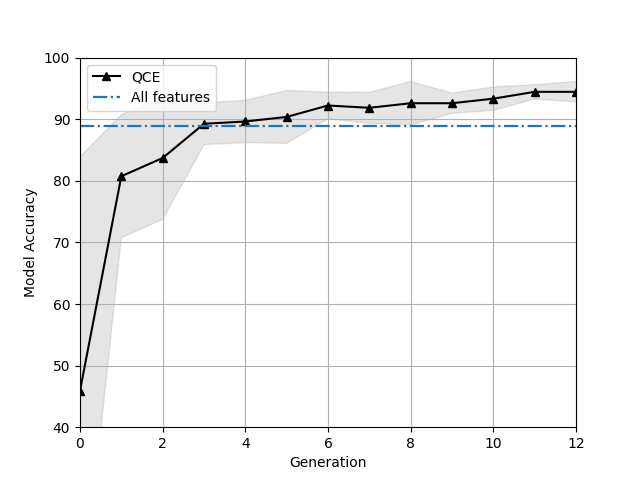}
    \caption{SVM test accuracy for the best individual at each generation. The dashed blue line represents the SVM test accuracy using all dimensions (features). The gray shadow area shows the standart deviation and the black line the mean, both for test accuracy.}
    \label{fig:bestres}
\end{figure}

\begin{figure}[h!]
    \centering    \includegraphics[width=7.5 cm]{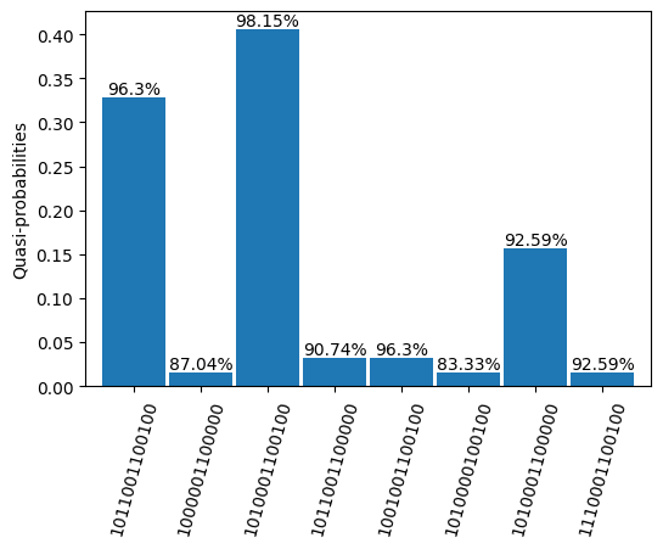}
    \caption{Final distribution (12th generation) over 64 measurements. The percentage at the top of each bar indicates the accuracy, $f(\mathbf{x})$, of SVC for the respective combination of features. The vector $\mathbf{x} = 1010001100100$ proves to be the best found combination of features for the model, whose test accuracy was $f(\mathbf{x})=98.15$).
    }
    \label{fig:finaldistribution}
\end{figure}

To carry out the experiments, a labeled data set of dimension $n=13$ (number of qubits) which uses chemical features to determine the origin of wines \cite{winedata} was used. We adopted the elitist scheme (6+1)EA with the following mutation probabilities: $50 \%$ to insert; $30 \%$ to modify; $10 \%$ delete and $10 \%$ swap. The total number of measurements performed on the quantum circuit was $m=64$, that is, $m \approx 5n$. Note that for $n=13$, $\text{dim}(|\psi\rangle) = 2^n = 8192$, so $m/2^n \approx 7.8e-3$. The metric used for $f(\mathbf{x})$ is the test accuracy of the Support Vector Classifier (SVC) model with a linear kernel function. The unseen labeled data used for testing is $20 \%$ of the total data. Fig. \ref{fig:bestres} shows the behavior of the quality of the solutions over the course of $K=12$ generations. The same experiment was performed $10$ times for statiscal analysis purposes.
In view of this, it was found that even for a small number, $m$, of measures and few generations of evolution of the circuit, the EQFS can find several combinations of features whose metric, $f(\mathbf{x})$, exceeds the case where the data set is used entirely. The best final distribution of the best individual among the $10$ experiments can be seen in Fig. \ref{fig:finaldistribution}. From this distribution, we obtained several different combinations of features with test accuracy superior to the case where all features are used ($88.8\%$).

\subsection{Model Evaluations}

\begin{figure}[h!]
    \centering    \includegraphics[width=7.5 cm]{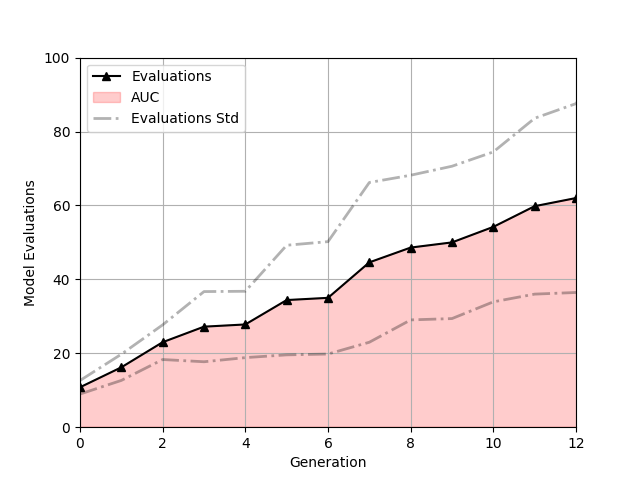}
    \caption{Total number of model evaluation at each generation. The Area Under the Curve (AUC) represents the total amount of model evaluation for all generations.}
    \label{fig:auc}
\end{figure}

The experimental evaluation of the model's performance was based on the total number of times that $f(\mathbf{x})$ is evaluated, since this is a critical point of the algorithm. In Fig. \ref{fig:auc} it can be seen that for a small number of generations, $K$, the total number of evaluations - for all $\lambda$ individuals - from $f(\mathbf{x})$ grows linearly as a function of the number of generations (with $m=64$).
The Area Under Curve (AUC) for the mean value of evaluations was $AUC=434$. This value shows that for a small constant $c \textless n$, the number of evaluations needed to find values equal to the quality of the presented experiments is $\mathcal{O}(c n^2)$.

\subsection{Quantum circuit depth}

\begin{figure}[h!]
    \centering    \includegraphics[width=4.5 cm]{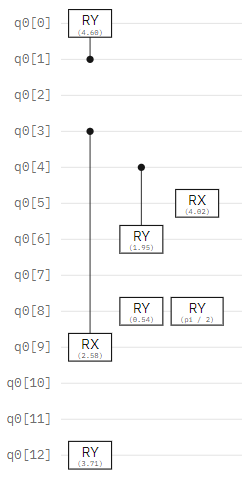}
    \caption{Final quantum circuit for a single run. The circuit depth is $p=3$, without any transpilation process for running on real quantum devices.}
    \label{fig:circuit}
\end{figure}

The depth of the quantum circuit for the number of generations $K$ obviously cannot exceed this value. Our experiments showed that the circuit depth - considering the base $B = \{RX, RY, RZ, RXX, RYY, RZZ\}$ - reached the average value of $p=3$, over all executions. The first run generated the circuit of Fig. \ref{fig:circuit}.

The Quantum Circuit Evolution heuristic has shown promise for solving combinatorial optimization problems in quantum computers, mainly due to the production of shallow quantum circuits. Fig. \ref{fig:circuit} showed that for the feature selection task this heuristic also produced circuits with low depth and a reduced level of entanglement (only 3 operations that generate entanglement between qubits), which further corroborates the suitability of this algorithm for NISQ computers, since two-qubit quantum gates have a lower fidelity than single-qubit gates.

Given the mutation probabilities chosen for the experiment, which are $30\%, 10\%, 10\%$ and $10\%$ for \textit{insert, delete, swap} and \textbf{modify}, respectively, we can see that the sum of the success rates of the \textit{delete, swap} and \textbf{modify} operations - which do not contribute to increasing the loop depth - is greater than that of the \textbf{insert} operation, since $p =3$ is $1/4$ of $k=12$. The worst case for loop depth would be $p=K$, for a scenario where \textbf{insert} would have a $100\%$ success rate.

\section{Conclusions} \label{IV}

In this study, we introduced a novel approach for feature selection based on Quantum Circuit Evolution (QCE) algorithm. Our results demonstrate that our procedure, Evolutionary Quantum Feature Selection (EQFS), can identify good feature combinations with a quadratic number of model evaluations. Additionally, we observed that the depth of the quantum circuits generated by EQFS was shallow and produced circuits with a small entanglement degree compared to their variational counterparts.

The effectiveness of our method highlights its potential to pave the way for the practical application of quantum computers in feature selection. With our findings, we hope to encourage further research in this area, as the potential impact of quantum computing on feature selection and other machine learning tasks could be significant. Our work contributes to the growing body of knowledge on quantum algorithms for machine learning and provides a promising new avenue for future research.

\section{ACKNOWLEDGEMENTS}
The authors would like to thank Banco Votorantim (BV) for providing resources for the project from which this article is derived, as well as for contributions and discussions throughout the work. Acknowledgements also to the Supercomputing Center for Industrial Innovation (CS2i), the Reference Center for Artificial Intelligence (CRIA), and the Latin American Quantum Computing Center (LAQCC), all from SENAI CIMATEC.

\bibliography{sample}

\end{document}